\documentclass[superscriptaddress,aps,journal=prl,preprint]{revtex4-2}
\usepackage{graphicx}
\usepackage{gensymb}
\usepackage{bm,amsmath,amssymb} 
\usepackage{color, soul} 
\usepackage{dcolumn}   
\usepackage{multirow}
\usepackage[colorlinks=true, linkcolor=blue, citecolor=blue, urlcolor=blue]{hyperref}

\newcommand{\EOP}{$\mathcal{E}_{3}$}
\newcommand{\POP}{$\mathcal{P}$}

\begin{document}

\title{Superlubric Motion of Wave-like Domain Walls in Sliding Ferroelectrics}

\author{Changming Ke}
\affiliation{Department of Physics, School of Science and Research Center for Industries of the Future, Westlake University, Hangzhou, 310030, China}
\affiliation{Institute of Natural Sciences, Westlake Institute for Advanced Study, Hangzhou, Zhejiang 310024, China}
\author{Fucai Liu}
\affiliation{School of Optoelectronic Science and Engineering, University of Electronic Science and Technology of China, Chengdu 611731, China.}
\affiliation{State Key Laboratory of Electronic Thin Films and Integrated Devices, University of Electronic Science and Technology of China, Chengdu 611731, China.}
\author{Shi Liu}
\email{liushi@westlake.edu.cn}
\affiliation{Department of Physics, School of Science and Research Center for Industries of the Future, Westlake University, Hangzhou, 310030, China}
\affiliation{Institute of Natural Sciences, Westlake Institute for Advanced Study, Hangzhou, Zhejiang 310024, China}

\date{\today}

\begin{abstract}{
Sliding ferroelectrics constructed from stacked nonpolar monolayers enable out-of-plane polarization in two dimensions with exceptional properties, including ultrafast switching speeds and fatigue-free behavior. However, the widely accepted switching mechanism, which posits synchronized long-distance in-plane translation of entire atomic layers driven by an out-of-plane electric field, has shown inconsistencies with experimental observations. We demonstrate that this spinodal decomposition-like homogeneous switching process violates Neumann's principle and  is unlikely to occur due to symmetry constraint.
Instead, symmetry-breaking domain walls (DWs) and the tensorial nature of Born effective charges are critical for polarization reversal, underscoring the quantum nature of sliding ferroelectrics. Using the Bernal-stacked $h$-BN bilayer as a model system, we discover that the coherent propagation of wide, wave-like domain walls is the key mechanism for ferroelectric switching. This mechanism  fundamentally differs from the layer-by-layer switching associated with narrow domain walls, which has been established for over sixty years in perovskite ferroelectrics. Moreover, these wave-like DWs exhibit superlubric dynamics, achieving ultrahigh velocities of approximately 4000 m/s at room temperature and displaying an anomalous cooling-promoted switching speed. The unexpected emergence of DW superlubricity in sliding ferroelectrics presents new avenues for enhancing key performance metrics and offers exciting opportunities for applications in cryogenic environments.
}
\end{abstract}
\maketitle
\newpage

Sliding ferroelectricity, achieved by stacking nonpolar monolayers with different rotational and translational configurations to induce switchable out-of-plane polarization (\POP), has significantly expanded the family of van der Waals (vdW) ferroelectrics in low-dimensional systems~\cite{Wu21pe2115703118, Wang23p542, Ji23p146801, Zhang22p25, ViznerStern24p}. 
Various sliding ferroelectrics consisting of nonpolar two-dimensional (2D) materials have been confirmed experimentally, including homobilayers of \textit{h}-BN~\cite{ViznerStern21p1462, Yasuda21p6549, Tsymbal21p6549} and transition metal dichalcogenides (TMDs)~\cite{Wang22p367, Weston22p390, Ko23p992}, as well as heterobilayers such as h-BN/graphene~\cite{Zheng20p71}, MoS$_2$/WS$_2$~\cite{Roge22p973}, and multilayer TMDs~\cite{Deb22p465, VanWinkle24p751, Meng22p7696}. The inherent 2D nature of the building blocks for sliding ferroelectrics provides numerous advantages typical of vdW materials, including atomic thickness and flexibility. Recently, 
ferroelectric field-effect transistors (FeFETs) based on sliding ferroelectricity in bilayer $h$-BN have demonstrated exceptional performance, featuring endurance exceeding 10$^{11}$ cycles and ultrafast switching on the nanosecond scale~\cite{Yasuda24peadp3575}. Likewise, sliding ferroelectricity in bilayer 3R-MoS$_2$ has also been utilized in a FeFET device, achieving antifatigue performance over a stress duration of 10$^5$ seconds~\cite{Bian24p57}.
Moreover, when composed of layers exhibiting additional functional properties such as magnetism or superconductivity, sliding ferroelectrics could effectively couple electrical polarization with other order parameters~\cite{Li17p6382, Wang23p542}, enabling the design of multifunctional materials that allow for dynamic tuning of electronic~\cite{Xiao20p1028, Jindal23p48}, photonic~\cite{Dong23p36,Yang22p469}, and magnetic properties~\cite{Huang18p544, Liu20p247601} under an applied electric field.

\begin{figure}[htb]
\centering
\includegraphics[scale=1, trim = 0 445 0 10, clip]{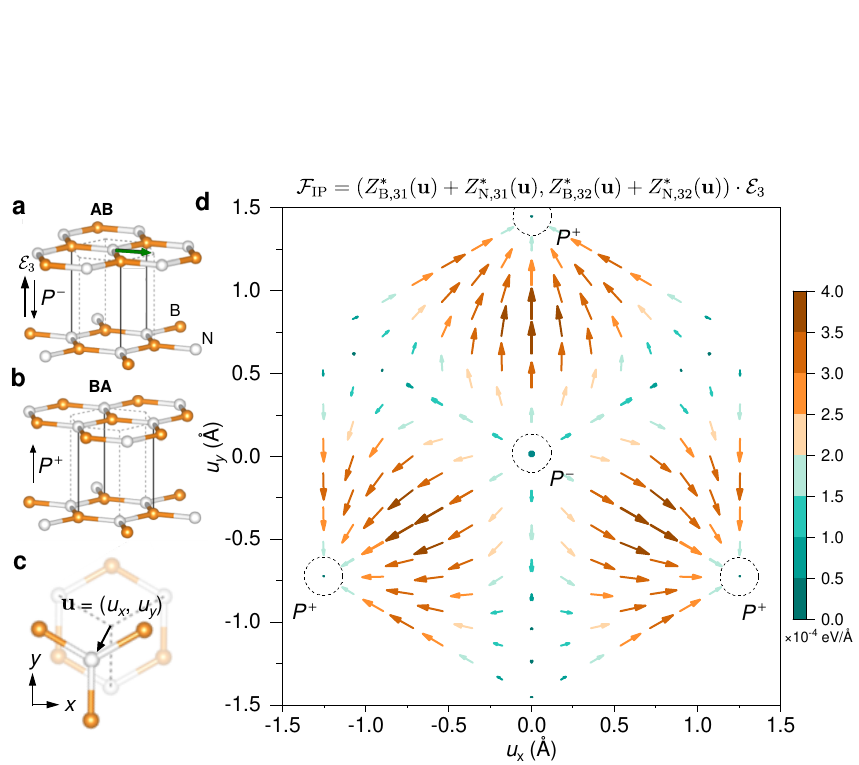}
 \caption{Tensorial nature of in-plane forces induced by an out-of-plane electric field in the Bernal-stacked $h$-BN bilayer. Atomic structures of (a) the AB-stacked configuration with downward polarization ($\mathcal{P}^-$) and (b) the BA-stacked configuration with upward polarization ($\mathcal{P}^+$). Orange and white spheres represent B and N atoms, respectively. An out-of-plane electric field ($\mathcal{E}_3$) needs to induce in-plane atomic forces to drive the transition from the AB to BA configuration. 
 (c) Top view of the bilayer illustrating the in-plane displacement vector $\mathbf{u}=(u_x, u_y)$, defined relative to the center of the hexagonal ring in the bottom layer. 
 (d) Distribution of the in-plane force $\mathcal{F}_{\rm IP}$ acting on the unit cell in the top layer as a function of displacement vector $\mathbf{u}$ under $\mathcal{E}_3=$ 0.3 V/nm. Each arrow represents the computed in-plane force vector for a specific bilayer configuration and is color-coded to indicate force magnitude. 
 The in-plane force is strictly zero in both ground states ($\mathcal{P}+$ and $\mathcal{P}^-$) due to the out-of-plane three-fold rotational symmetry (C$_3$). Unit cells with C$_3$ symmetry broken will experience in-plane forces arising from non-zero off-diagonal BEC components, $Z_{\kappa,31}^*$ and $Z_{\kappa,32}^*$ with $\kappa=$~B, N.
}
\label{fig_bec}
\end{figure}

The functionality of a ferroelectric hinges on the process of electric field-driven polarization reversal. 
In sliding ferroelectrics, the widely accepted mechanism involves long-distance lateral displacement of the two monolayers relative to each other, as exemplified by the Bernal-stacked \textit{h}-BN bilayer in Fig.\ref{fig_bec}a-b. Specifically, an upward out-of-plane electric field (\EOP) is assumed to induce opposing in-plane atomic motions in the top and bottom layers, transforming the stacking configuration from AB to BA and flipping the direction of \POP~\cite{Li17p6382}. This mechanism has gained support from zero-Kelvin density functional theory (DFT) calculations at the unit cell level, which predict sliding barriers of just a few meV per unit cell.  It is important to note that in these theoretical investigations, the sliding pathway is constructed by manually dragging the layers, without explicitly accounting for the influence of the out-of-plane electric field~\cite{Zhong23pe16832, Wang23p035426}. These low barriers have been invoked to explain the observed ultrafast switching speed and fatigue-resistant behavior of sliding ferroelectrics. 

However, careful analysis of this prevailing mechanism reveals several fundamental inconsistencies. First, 
both AB and BA-stacked configurations belong to the $P3m1$ space group, exhibiting out-of-plane three-fold rotational symmetry (C$_3$). According to Neumann's principle~\cite{Neumann85book}, applying \EOP~along this rotation axis \textit{cannot} generate a \textit{net} in-plane force that would break this symmetry, a conclusion confirmed by our finite-field DFT calculations (see discussions below).
Second, the assumed synchronized translation of an entire layer of atoms resembles spinodal decomposition~\cite{Binder87p783}, where all dipoles flip simultaneously without nucleating oppositely polarized domains. 
It can be readily demonstrated that, for any macroscopic size of the sample, this homogeneous process 
is associated with an impractically high energy barrier, many orders of magnitude greater than the thermal energy, making the switching probability vanishingly small (see Supplementary Section II). 
Finally, the collective nature of the proposed mechanism implies that even minor variations in the local environment, such as temperature fluctuations or the presence of local defects, could impede the switching process, which is inconsistent with the observed fatigue-free characteristics.   

Here, we present a general theory of polarization switching in sliding ferroelectrics that addresses the experiment-theory conundrums discussed above.  Using the $h$-BN bilayer as a model system, we demonstrate that the tensorial nature of Born effective charges (BECs), an aspect largely overlooked in previous investigations, is essential for the emergence of \EOP-induced in-plane driving forces. Importantly, only atoms at the domain walls (DWs), which break the C$_3$ symmetry, possess non-zero off-diagonal BEC elements, enabling the generation of net in-plane forces that drive DW motion. Unlike conventional ferroelectrics, where atoms in both domains and DWs respond to external electric fields, sliding ferroelectrics restrict switching to atoms near DWs, while those within domains remain inert. Consequently, polarization reversal in sliding ferroelectrics is governed by DW motion, involving only local atomic displacements rather than global interlayer shifts. These distinctive characteristics are confirmed by deep potential molecular dynamics (MD) simulations combined with a local atomic environment-dependent BEC model that captures dynamic charge transfer during DW motion. Notably, we discover that DWs in the $h$-BN bilayer propagate via wave-like coherent atomic displacements, contrasting sharply with the classic layer-by-layer mechanism for DWs in perovskite ferroelectrics. Moreover, field-driven DW dynamics can be described as superlubric motion featuring a zero motion barrier, achieving ultrahigh velocities of $\approx$4000 m/s at room temperature and exhibiting anomalous  cooling-enhanced switching speed. The unexpected emergence of DW superlubricity in sliding ferroelectrics opens up opportunities for optimizing coercive fields and switching speeds in these systems, as well as for designing advanced electronics that function at cryogenic conditions.

We first establish that both the tensorial nature of BECs and the breaking of C$_3$ symmetry underpin polarization reversal in sliding ferroelectrics. 
In the AB-stacked \textit{h}-BN bilayer, the boron (B) atom in the upper layer is positioned directly above the nitrogen (N) atom in the lower layer, resulting in charge transfer that generates a polarization along the $-z$ direction ($\mathcal{P}^-$), as illustrated in Fig.~\ref{fig_bec}a. In a simple point-charge model, where each atom is assigned a fixed charge, applying \EOP~generates forces solely in the out-of-plane direction, rendering in-plane atomic motion impossible. In contrast, the BEC directly reflects the force on atom $\kappa$ in direction $j$ ($\mathcal{F}_{\kappa,j}$) induced by a macroscopic electric field in direction $i$ ($\mathcal{E}_{i}$) through the relationship $Z_{\kappa, ij}^*=\frac{\partial \mathcal{F}_{\kappa,j}}{\partial \mathcal{E}_{i}}$. The off-diagonal elements of the BEC tensor ($Z^*_{31}$ and $Z^*_{32}$) are critical for generating in-plane atomic forces. However, the presence of C$_3$ symmetry cancels these in-plane forces in both AB and BA configurations. For each unit cell, the net in-plane force can be approximated as $\mathcal{F}_{\rm IP} = (Z^*_{\rm B,31}+Z^*_{\rm N, 31}, Z^*_{\rm B,32}+Z^*_{\rm N, 32})\cdot \mathcal{E}_{3}$. Using BEC values computed with DFT (see Table S1), we find that the magnitude of $\mathcal{F}_{\rm IF}$ is strictly zero in AB-stacked and BA-stacked bilayers (see Fig.~\ref{fig_bec}d). These results indicate that applying \EOP~along the polarization direction of a single-domain state in sliding ferroelectrics, surprisingly, produces no driving force for in-plane atomic motion. 

We further map out $\mathcal{F}_{\rm IF}$ in the top layer as a function of the in-plane displacement vector for N and B atoms, $\mathbf{u}=(u_x, u_y)$, defined relative to the center of the hexagonal ring in the bottom layer (see Fig.~\ref{fig_bec}c), using BEC values
computed individually at each $\mathbf{u}$. As shown in Fig.~\ref{fig_bec}d, the magnitude of $\mathcal{F}_{\rm IF}$ correlates with the extent to which C$_3$ symmetry is broken. The pronounced variation in $\mathcal{F}_{\rm IF}$ highlights the strong dependence of off-diagonal BEC elements on the local atomic environment, reflecting substantial charge transfer during sliding. With these DFT results, we derive a set of analytical functions, $\mathcal{Z}_{3j}(u_x, u_y)$, which accurately predict unit-cell-averaged BEC tensors, $(Z^*_{\rm B,31}+Z^*_{\rm N, 31}, Z^*_{\rm B,32}+Z^*_{\rm N, 32}, Z^*_{\rm B,33}+Z^*_{\rm N, 33})$, from local sliding displacements (see Supplementary Section III).

The unswitchable nature of single domains in sliding ferroelectrics strongly suggests that the experimentally observed polarization reversal originates from structural units where C$_3$ symmetry is broken. The DWs separating AB-stacked and BA-stacked regions naturally serve as C$_3$-symmetry-breaking interfaces. We optimize the structure of $\Sigma_0$-type 180\degree~DWs (see classification of DW types in Supplementary Section IV) in the $h$-BN bilayer with a first-principles-derived machine learning force field (see Computational Methods in Supplementary Section I). As illustrated in Fig.~\ref{fig_addf}a, the $\Sigma_0$ wall extends parallel to the displacement vector $\mathbf{u}$, separating the $\mathcal{P}^+$ domain ($u_x=0$, $u_y=u_0$) and the $\mathcal{P}^-$ domain ($u_x=0$, $u_y=0$). Notably, this wall exhibits nearly zero out-of-plane buckling, whereas other types of walls induce ripples in the bilayer (Fig.~S6). 
The unit-cell-resolved local polarization ($p_{\rm op}$) is gauged by the structural parameter $\lambda = 2u_y/u_0-1$, which evolves from $+1$ to $-1$ across the wall. 
In stark contrast to the narrow 180\degree~DWs in perovskite ferroelectrics, which are only a few unit cells wide ($<1$ nm) ~\cite{Liu16p360}, the optimized DW structure in the $h$-BN bilayer exhibits a much larger width ($w$) of approximately 10 nm (Fig.~\ref{fig_addf}a-c). Using the $\mathcal{Z}_{3j}(u_x, u_y)$ functions for local BEC predictions, we estimate unit-cell-averaged in-plane forces in the top layer under \EOP~= 0.3 V/nm, with the results depicted in Fig.~\ref{fig_addf}d. Remarkably, only the unit cells near the DW experience collective in-plane forces along the $y$-axis, as C$_3$ symmetry is broken in this region. In contrast, unit cells farther away from the DW do not acquire any in-plane forces.
The polarization reversal mechanism in sliding ferroelectrics becomes evident: unit cells at the DW experience \EOP-induced in-plane forces along the $y$ axis, which alter the magnitude of in-plane $u_y$ displacements, leading to DW motion. 

We investigate DW dynamics in the $h$-BN bilayer using finite-field MD simulations, with $\mathcal{Z}_{3j}(\mathbf{u})$ functions implemented to capture dynamic charge transfer and the evolving field-induced forces in real time. 
Our approach fundamentally differs from previous MD simulations of DWs in sliding ferroelectrics~\cite{He24p119416}, which applied a uniform force to each atom, regardless of its charge or local environment. 
The evolution of the domain structures is illustrated in Fig.~\ref{fig_addf}e. 
Simulations considering only $Z_{33}^*$ show that the $\Sigma_0$ wall experiences no net lateral displacement after 1 nanosecond under a high field strength of 3 V/nm.
In contrast, simulations that correctly include  $Z_{31}^*$ and $Z_{32}^*$ reveal DW motions under \EOP~= 0.3~V/nm. Interestingly, results account for all elements of $Z_{3i}^*$ are similar to those incorporating only the off-diagonal components, confirming that these off-diagonal BEC elements play a dominant role in polarization reversal.  Considering the tensorial nature of BECs is thus necessary to fully capture the quantum origin of sliding ferroelectrics. Additionally, once the bilayer transitions to a single-domain state after DW motion, applying an opposite field fails to generate new DWs for polarization switching. Using the same approach, we reproduce the \EOP-induced pattern evolution of triangular domains in twisted \textit{h}-BN with a twist angle of 0.336\degree~exhibiting moir\'e ferroelectricty (Fig.~\ref{fig_addf}f). The moving DWs behave like tightly-strung bowstrings, resulting in distinctive curved triangular domains that are consistent with experimental observations~\cite{Ko23p992, Weston22p390}. The curvature of the DWs arises naturally from immobile corners protected by C$_3$ symmetry that effectively nullifies in-plane forces. 

\begin{figure}[htb]
\centering
\includegraphics[scale=0.7, trim = 85 123 30 5, clip]{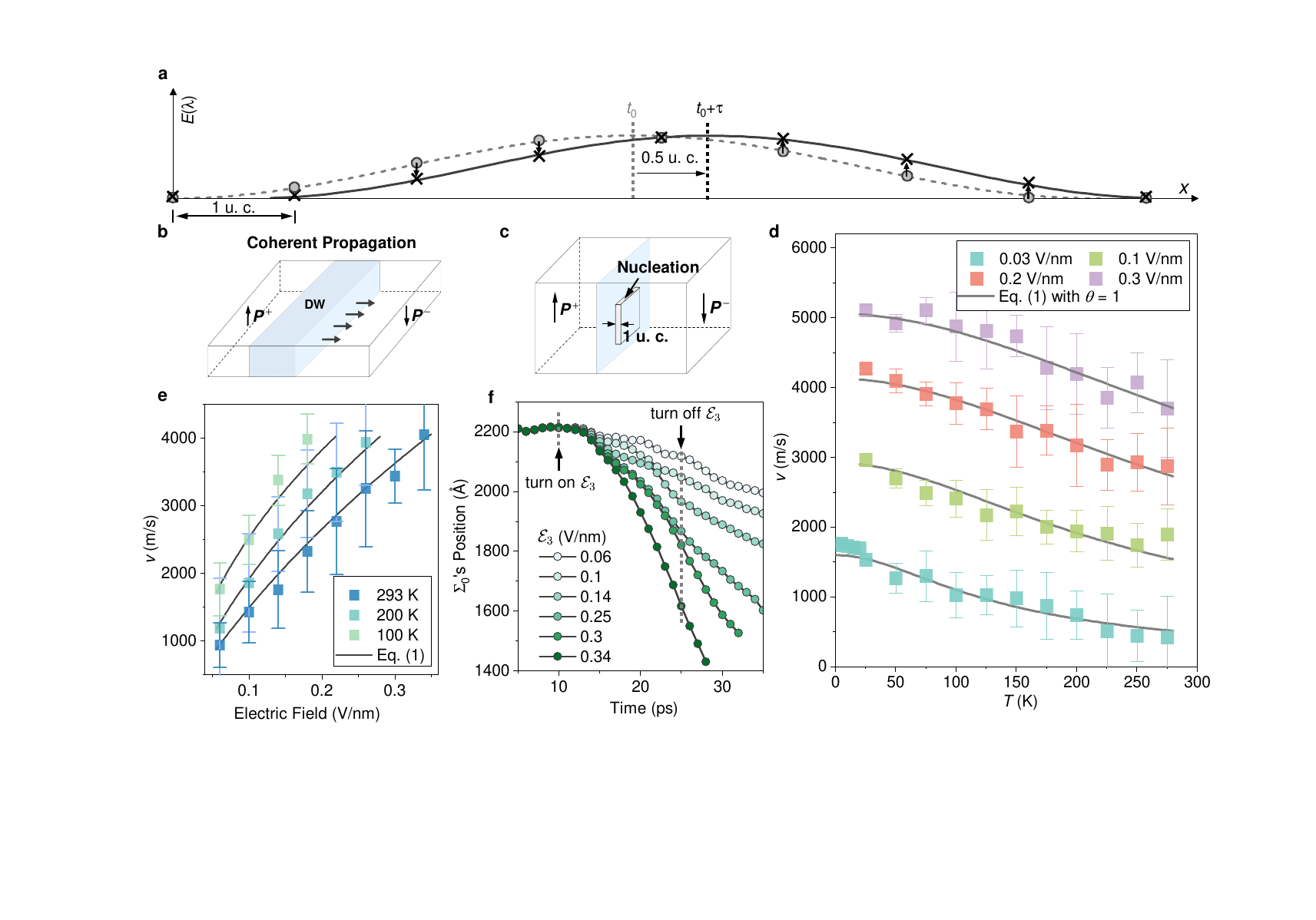}
  \caption{Superlubric motion of wave-like domain walls. (a) Evolution of the local energy profile $E({\lambda})$ for a wide DW translating by half of a unit cell (0.5 u.c.). The height of the symbols (circles and crosses) corresponds to the local energy $E({\lambda})$ of the unit cell, characterized by the structural order parameter $\lambda=2u_y/u_0-1$. During the motion, unit cells with decreasing local energies balance those with increasing local energies, resulting in nearly zero motion barrier. Schematic representations of (b) coherent propagation of a wide domain wall in sliding ferroelectrics and (c) layer-by-layer switching through nucleation-and-growth in perovskite ferroelectrics. (d) Velocity of the  $\Sigma_0$ DW under varying temperatures and electric fields. (e) Cooling-promoted DW motion. (f)  Inertial response of moving DWs  following the removal of \EOP. The walls maintain nearly the same velocity at the moment of field removal, indicating zero damping and a superlubric effect.}
   \label{fig_superl}
 \end{figure}

Our MD simulations reveal that atoms within each unit cell simply need to undergo small displacements to enable long-distance travel of DWs. As illustrated in Fig.~\ref{fig_addf}e, for a wall that travels a distance of $l$, the maximum local displacement is given by $\frac{l}{w}u_0$. For instance, if the wall moves 35~\AA~along the $x$ direction, the maximum in-plane local atomic movement is limited to just 0.5~\AA~due to the wall's substantial width. The displacement of the DW is accomplished through the collective and coordinated small in-plane shifts of atoms at the wall, resembling the propagation of a wave packet.
Figure~\ref{fig_superl}a sketches the evolution of the local energy profile $E(\lambda)$ for a wall that moves by half of a unit cell. A key feature is the balance between unit cells with decreasing local energies and those with increasing local energies, resulting in nearly zero motion barrier. Such wave-like ``coherent propagation" (Fig~\ref{fig_superl}b) is markedly different from 
the layer-by-layer switching mechanism pioneered by Miller and
Weinreich for sideway motion of narrow DWs in perovskite ferroelectrics (Fig.~\ref{fig_superl}c), which involves the formation and growth of a 2D nucleus at the wall~\cite{Miller60p1460}. 
The wave-like DW motion in the $h$-BN bilayer exhibits near-barrierless behavior, leading to ultrahigh DW velocities ($v$), as confirmed by our finite-field MD simulations. For instance, at 293~K and \EOP = 0.3 V/nm, the DW achieves a speed of $\approx3000$ m/s, agreeing reasonably well with estimated experimental value~\cite{Yasuda24peadp3575}.

Importantly, the observation that a flat bilayer enables  barrierless motion of $\Sigma_0$ walls closely resembles the phenomenon of structural superlubricity~\cite{Wang24p011002}, where static friction between theoretically infinite crystal contacts is effectively zero at low temperatures. Under steady-state sliding conditions, the kinetic friction $\mathcal{F}_k$ between 2D superlubirc interfaces can be expressed as $\mathcal{F}_k \propto v\sqrt{c_{_h} h^4+c_{_T}(k_BT)^2}$~\cite{Wange23p105396}, where $v$ is the sliding velocity, $h$ is the corrugation height, and $c_{_h}$ and $c_{_T}$ are material-specific parameters. Assuming that the electrical driving force balances $\mathcal{F}_k$, we establish heuristically the temperature and field dependence of the $\Sigma_0$ wall's velocity as
\begin{equation}
    v\propto \frac{\mathcal{E}}{\sqrt{\gamma\mathcal{E}^\theta + c_{_T} (k_BT)^2}},
\end{equation}
where $\gamma \mathcal{E}^\theta$ accounts for the influence of the electric field on $h$. Despite the simplicity of the argument, Equation (1) effectively describes velocity data from MD simulations across a wide range of temperatures and out-of-plane fields (Fig.~\ref{fig_superl}e-f) with $\theta=1$, supporting its physical basis. Notably, DW velocity increase with decreasing temperature, which can be attributed to reduced friction from suppressed thermal corrugation at lower temperatures, similar to the facilitated sliding observed at lower temperatures in structurally superlubric 2D material interfaces~\cite{Guerra10p634}.
This trend stands in stark contrast to the creep behavior of DWs in perovskite ferroelectrics, where $v$ increases with rising temperature~\cite{Jo09p045701, Tybell02p97601, Liu16p360, Bai24p26103}. The superlubric nature of the $\Sigma_0$ wall is further corroborated by its strong inertial response after the field is removed. As shown in Fig.~\ref{fig_superl}g, the wall continues to move at nearly the same velocity even after the driving force is turned off, which differs significantly from the absence of inertial motion observed in conventional perovskite ferroelectrics~\cite{Liu13p232907}. Additionally, it does not escape our observation that the DW velocity saturates as the temperature approaches absolute zero. We suggests that the dynamics of DWs in sliding ferroelectrics at ultralow temperatures may be described by a Lorentz-invariant sine-Gordon equation, wherein a maximum characteristic velocity emerges, playing a role analogous to the speed of light in the theory of special relativity~\cite{Caretta20p1438}.

\begin{figure}[htb]
\centering
\includegraphics[scale=0.8, trim = 0 18 10 20, clip]{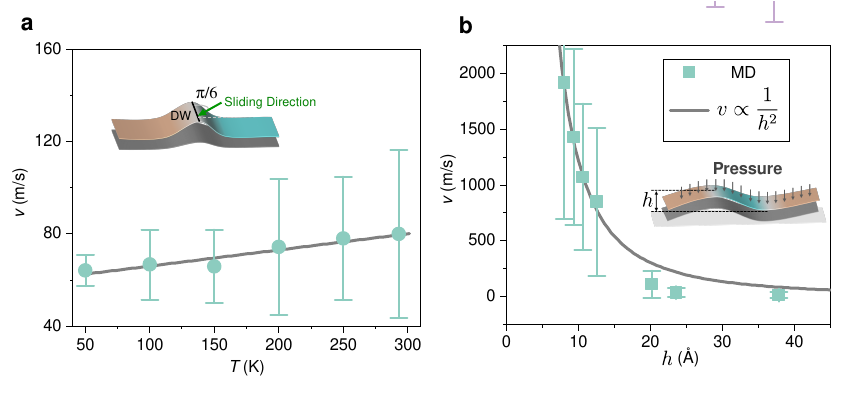}
 \caption{ Creep-to-superlubric transition of DW motions. (a) Temperature dependence of DW velocity for $\Sigma_{\pi/6}$-type walls exhibiting out-of-plane buckling. The velocity increases with rising temperature. (b) DW velocity as a function of out-of-plane buckling height ($h$). Ultrahigh domain wall speeds at 5 V/nm are achieved by applying out-of-plane compressive stress that reduces $h$.}
  \label{fig_pres}
\end{figure}

Finally, we demonstrate that other types of DWs with substantial buckling can be made sperlubric-like by applying an out-of-plane compressive stress that flattens the bilayer. For example, the extension direction of a $\Sigma_{\pi/6}$-type 180\degree~DW forms an angle of $\pi/6$ with the local displacement vector, and its ground-state structure has a substantial bucking of 38~\AA. 
The temperature dependence of the DW velocity for the $\Sigma_{\pi/6}$ wall is similar to DWs in conventional ferroelectrics, where $v$ correlates positively with $T$, as shown in Fig~\ref{fig_pres}a. 
By applying an out-of-plane compressive stress to reduce the corrugation, the velocity at \EOP~=~5 V/nm increases rapidly as $h$ becomes smaller. We find a relationship of $v \propto 1/h^2$ (Fig.~\ref{fig_pres}b), which is comparable to the corrugation-dominated kinetic friction equation, $\mathcal{F}_k \propto v h^2 $~\cite{Wang24p011002}.

In summary, the DW-driven polarization reversal in sliding ferroelectrics presented here is fundamentally distinct from the widely accepted spinodal decomposition-like switching mechanism, which requires the global translation of monolayers. The quasi-1D DW, with its large width, facilitates wave-like coherent propagation, setting it apart from the classic layer-by-layer switching mechanism established sixty years ago for lateral motion of 2D DWs. The presence of a DW is essential for switching in sliding ferroelectrics, as a single-domain state cannot nucleate new domains due to symmetry protection. This feature can be leveraged to achieve deterministic control of polar states as DWs are the only structural units responsive to electrical stimuli, avoiding the stochastic switching behavior commonly associated with the formation of new domains and DWs through nucleation in perovskite ferroelectrics. We believe that sliding ferroelectrics hold great promise for the deterministic and repeatable production of multi-state polarization for neuromorphic applications. The superlubric motion of wave-like DWs, as demonstrated in $\Sigma_0$ walls of the $h$-BN bilayer, explains the ultrafast switching speeds and fatigue-free behavior observed experimentally. It also indicates the potential of ``ferro-tribodynamics": applying tribological principles to optimize the kinetics of dipolar excitations. The anomalous  cooling-promoted switching speed highlights the promise of sliding ferroelectrics for high-performance nanoelectronics in cryogenic environments, while offering a platform to study Lorentz-invariant ferroelectric DW dynamics with relativistic-like kinematics.~\cite{Caretta20p1438, Yasuda24peadp3575}.
\\

{\bf{Acknowledgments}} C.K. and S.L. acknowledges the supports from Natural Science Foundation of China (52002335). F.L is supported by the National Key Research \& Development Program (2020YFA0309200) and National Natural Science Foundation of China (92477115). The computational resource is provided by Westlake HPC Center.

{\bf{Competing Interests}} The authors declare no competing financial or non-financial interests.

{\bf{Data Availability}} The data that support the findings of this study are included in this article and are available from the corresponding author upon reasonable request.

{\bf{Author Contributions}} S.L. led the project. C.K. performed simulations and data analysis and discovered superlubric motion of domain walls. All authors contributed to the discussion and the manuscript preparation.

\bibliography{SL}

\newpage

\begin{figure}[htb]
\centering
\includegraphics[scale=0.6, trim = 10 218 0 0, clip]{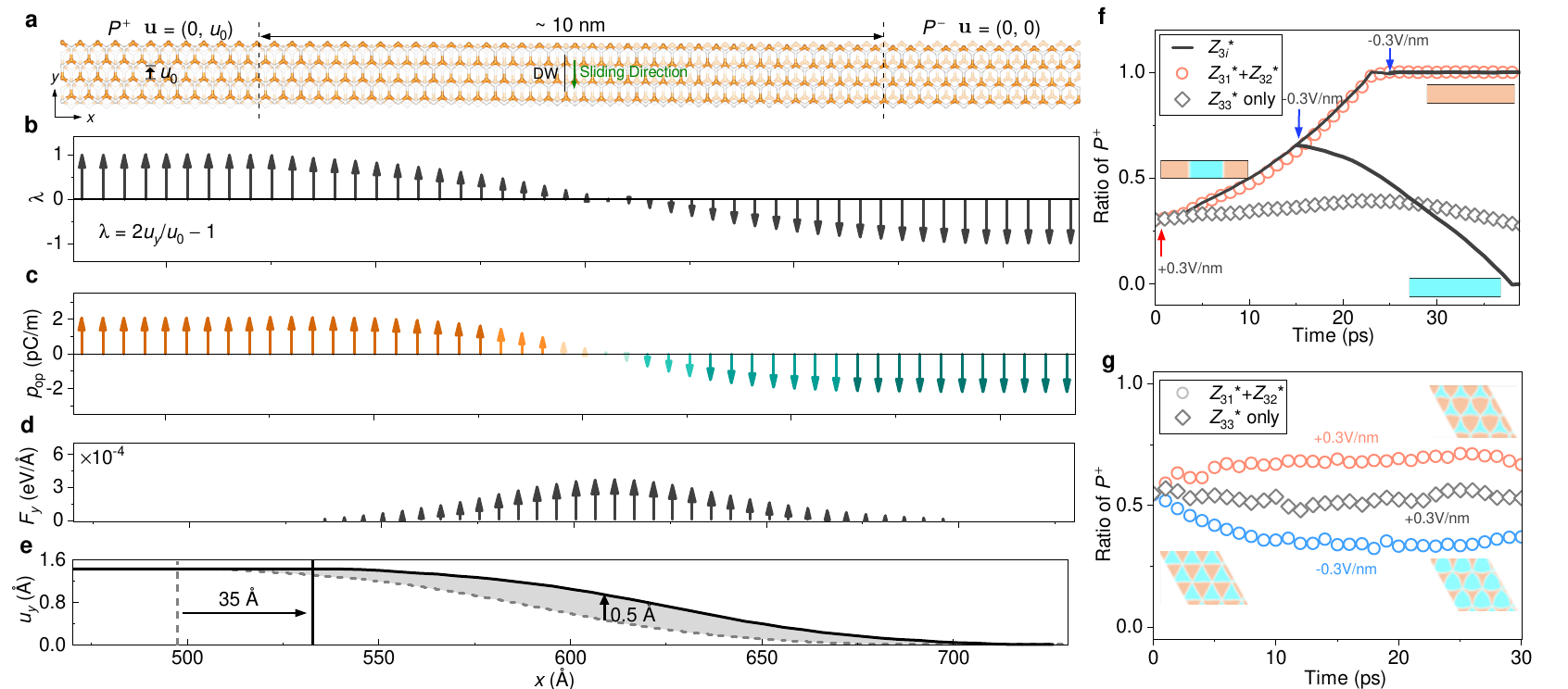}
 \caption{MD simulations of domain walls in sliding ferroelectrics with dynamic Born effective charges. (a) Schematic of a $\Sigma_0$-type wall separating the $\mathcal{P}^-$ domain with $\mathbf{u}=(0,0)$ and the $\mathcal{P}^+$ domain with with $\mathbf{u}=(0,u_0)$. The wall extends along the $y$ axis and is parallel to the displacement vector $\mathbf{u}$. Profiles of (b) the structural order parameter $\lambda$ and (c) unit-cell-resolved local polarization ($p_{\rm op}$) reveal a wide DW with a thickness of $\approx10$~nm. The spacing between neighboring arrows is two unit cells for visual clarity. (d) Profile of unit-cell-averaged in-plane forces ($\mathcal{F}_y$) in the top layer induced by $\mathcal{E}_3=0.3$ V/nm. Only unit cells near the DW experience collective in-plane forces along the $y$-axis. (e) Geometric relationship between DW propagation distance and local atomic displacement; the wall moves by 35~\AA~ along the $x$-axis, whereas the maximum local atomic displacement is 0.5~\AA~along the $y$-axis. (f) Evolution of domain structures with $\Sigma_0$ walls using finite-field MD simulations that consider various components of the Born effective charges. Simulations that consider off-diagonal BEC elements ($Z_{31}^*$ and $Z_{32}^*$) correctly predict \EOP-driven DW motions. At 15~ps, applying a downward field of $-0.3$ V/nm reverses the switching process as expected. At 25~ps, the system has transitioned to a single-domain state and becomes unswitchable under \EOP~= $-0.3$ V/nm. 
 (g) Pattern evolution of triangular domains in twisted \textit{h}-BN with a twist angle of 0.336\degree~obtained from MD simulations considering dynamic Born effective charges.
  }
  \label{fig_addf}
\end{figure}

\clearpage
\newpage

\clearpage
\newpage

\end{document}